\documentclass[lineno,authoryear]{FLO_v1}%

\usepackage{graphicx}
\usepackage{upgreek}
\usepackage{multicol,multirow}
\usepackage{amsmath,amssymb,amsfonts}
\usepackage{mathrsfs}
\usepackage{amsthm}
\usepackage[figuresright]{rotating}
\usepackage{appendix}
\usepackage[authoryear]{natbib}
\usepackage{ifpdf}
\usepackage[T1]{fontenc}
\usepackage{newtxtext}
\usepackage{newtxmath}
\usepackage{textcomp}
\usepackage{xcolor}
\usepackage{color} 

\usepackage[colorlinks,allcolors=blue]{hyperref}
\definecolor{jourcolor}{cmyk}{1,0.57,0.01,0.38}
\hypersetup{
    colorlinks,%
    citecolor=jourcolor,%
    filecolor=jourcolor,%
    linkcolor=jourcolor,%
    urlcolor=jourcolor
}

\theoremstyle{definition}

\articletype{RESEARCH ARTICLE}

\DOI{xx.xxxx/flo.2022.xx}

\Year{2022}

\Vol{x}

\Price{}

\art-id{FLO2000049}

\citearticle{Zhong, J., Liu, S. \& Sun, C.}

\begin{document}

\title[On the thermal effect of porous material in porous media Rayleigh-Bénard convection]{On the thermal effect of porous material in porous media Rayleigh-Bénard convection}

\author[]{Jun Zhong$^{1}$, Shuang Liu$^{1,2}$ and Chao Sun$^{1,3\ast}${{\href{https://orcid.org/0000-0002-0930-6343}{\includegraphics{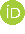}}}}}

\address[1]{Center for Combustion Energy, Key Laboratory for Thermal Science and Power Engineering of Ministry of Education, and Department of Energy and Power Engineering, Tsinghua University, 100084 Beijing, China}
\address[2]{Yau Mathematical Sciences Center, Tsinghua University, 100084 Beijing, China}
\address[3]{Department of Engineering Mechanics, School of Aerospace Engineering, Tsinghua	University, 100084 Beijing, China}

\corres{*}{Corresponding author. E-mail:
\emaillink{chaosun@tsinghua.edu.cn}}


\date{\textbf{Received:} XX 2022; \textbf{Revised:} XX XX 2022; \textbf{Accepted:} XX XX 2022}

\abstract{We perform a two-dimensional numerical study on the thermal effect of porous media on global heat transport and flow structure in Rayleigh-Bénard (RB) convection, focusing on the role of thermal conductivity $\lambda$ of porous media, which ranges from $0.1$ to $50$ relative to the fluid. The simulation is carried out in a square RB cell with the Rayleigh number $Ra$ ranging from $10^7$ to $10^9$ and the Prandtl number $Pr$ fixed at $4.3$. The porosity of the system is fixed at $\phi=0.812$, with the porous media modeled by a set of randomly displayed circular obstacles. For a fixed $Ra$, the increase of conductivity shows a small effect on the total heat transfer, slightly depressing the Nusselt number. The limited influence comes from the small number of obstacles contacting with thermal plumes in the system as well as the counteraction of the increased plume area and the depressed plume strength. The study shows that the global heat transfer is insensitive to the conduction effect of separated porous media in the bulk region, which may have implications for industrial designs.}

\maketitle
\begin{boxtext}

\textbf{\mathversion{bold}Impact Statement}

Convective flows in porous media occur in a wide variety of applications, and the presence of the porous structures has important effects on the flow structure and heat transfer efficiency in porous media convection.  
Both momentum transfer and heat exchange exist between porous media and the fluid, while how they influence the total heat transfer has not been thoroughly revealed so far. Here, we perform pore-scaled direct numerical simulations to study effect of thermal conductivity of porous media on heat transfer efficiency in Rayleigh-Bénard convection. It is surprisingly found that the global heat transfer efficiency only weakly depends on the thermal conductivity of the porous material. For real-world applications, our findings give implications for industrial designs of convection systems with the presence of obstacles; for the fundamental research, this work advances the understanding of the effect of porous media on natural convection.
\end{boxtext}

\section{Introduction}

Natural convection, as a typical fluid motion induced by the inhomogeneous density distribution of fluid, is a common phenomenon in nature and industrial processes. Rayleigh-Bénard (RB) convection is one of its general paradigms. In RB cells, fluid is heated at the lower plate with a constant high temperature $T_1$ and cooled at the upper plate with a constant low temperature $T_2$. Due to the temperature difference, the thermal expansion of the fluid generates an unstable density profile and drives the convection under gravity. Fluid motion in RB cells, especially turbulent RB convection, is one of the classical problems of fluid dynamics and has been studied extensively in the last several decades \citep{ahlers2009heat,lohse2010small,chilla2012new,xia2013current,Verma2018Physics,jiang_supergravitational_2020,wang_effects_2022}. A large-scale circulation (LSC) is formed in the system and 
takes thermal plumes from one horizontal boundary layer to the other one, strongly mixing the temperature field and enhancing heat transfer \citep{niemela2001wind,sun2005three,xi2004laminar,zhou2007morphological}. The interplay of LSC, thermal plumes, and boundary layers has been one of the central research issues for the study of turbulent thermal convection \citep{ahlers2009heat}.

When the porous media participate in convection, the phenomena become even richer \citep{pirozzoli_towards_2021,de_paoli_dissolution_2017,gasow_effects_2020,gasow_macroscopic_2021,gopalakrishnan_instability_2020}. In general, the large-scale convection is restrained, and the LSC even can disappear under a certain porosity. However, though the strength of the convection motion is depressed, the experiments and numerical studies show that the global heat transfer may even be enhanced due to the increase in the coherence of the flow. This phenomenon has been found in porous media convection \citep{liu2020rayleigh,liu2021lagrangian,ataei2019flow}, confined RB systems, and rotating RB systems \citep{chong2015condensation,chong2017confined,zhong2009prandtl,chong2018effect}. Liu {\it et al.} \citep{liu2020rayleigh} analyzed two-dimensional (2D) RB convection in porous media using direct numerical simulations, revealing that porous media have two competing effects: enhancing heat transfer by making the flow more orderly and suppressing heat transfer by reducing the flow strength. The effect of enhancing flow coherence is more pronounced when the typical pore scale is larger than the thickness of the thermal boundary layer, so in this situation, total heat transfer is increased as compared to the classical RB system. These findings are also confirmed in experiments \citep{ataei2019flow}.

In most of those studies, the thermal properties of the porous media are assumed to be identical to those of the working fluid. It is known that the porous media also contribute to heat transfer by thermal conduction, which is conspicuous in Porous Media Combustion \citep{kamal2006combustion,ferguson2021pore}. The porous media can absorb heat from the fluid and release heat to the fluid under different temperature differences, modulating the temperature fluctuations, and consequently the flow dynamics will be influenced due to the coupling effects of the temperature and velocity fields. In most previous studies, the thermal properties of porous media are set as the same as those of the fluid for simplification. In practice, the porous media can be plastic, metallic, or be made from other kinds of materials, with a large variation in thermal properties, which may result in a great difference in total heat transfer and flow behaviors of the system. This paper aims to investigate the thermal effect of porous media on the global heat transfer and flow structure in the RB system with porous media.

In this work, we conduct a 2D numerical study on the porous-media RB convection, constructing the porous media by a set of randomly displayed but contactless circular obstacles. We consider an important thermal property of porous media — thermal conductivity — and investigate its influence on the heat transfer and flow structure of RB system, attempting to expand our understanding of the effect of porous media on heat transfer in thermal convection system. The thermal conductivity of porous media relative to the fluid varies from $0.1$ (plastic) to $50$ (metal) in the study to reconcile the numerical results with reality. 

The rest of the paper is organized as follows: the establishment of the numerical model is introduced in section \ref{sec2}, while the main results are discussed in section \ref{sec3}. Finally, conclusions are presented in section \ref{sec4}.

\section{Numerical Model\label{sec2}}

\begin{figure}
	\centering
	\includegraphics[width=0.3\linewidth]{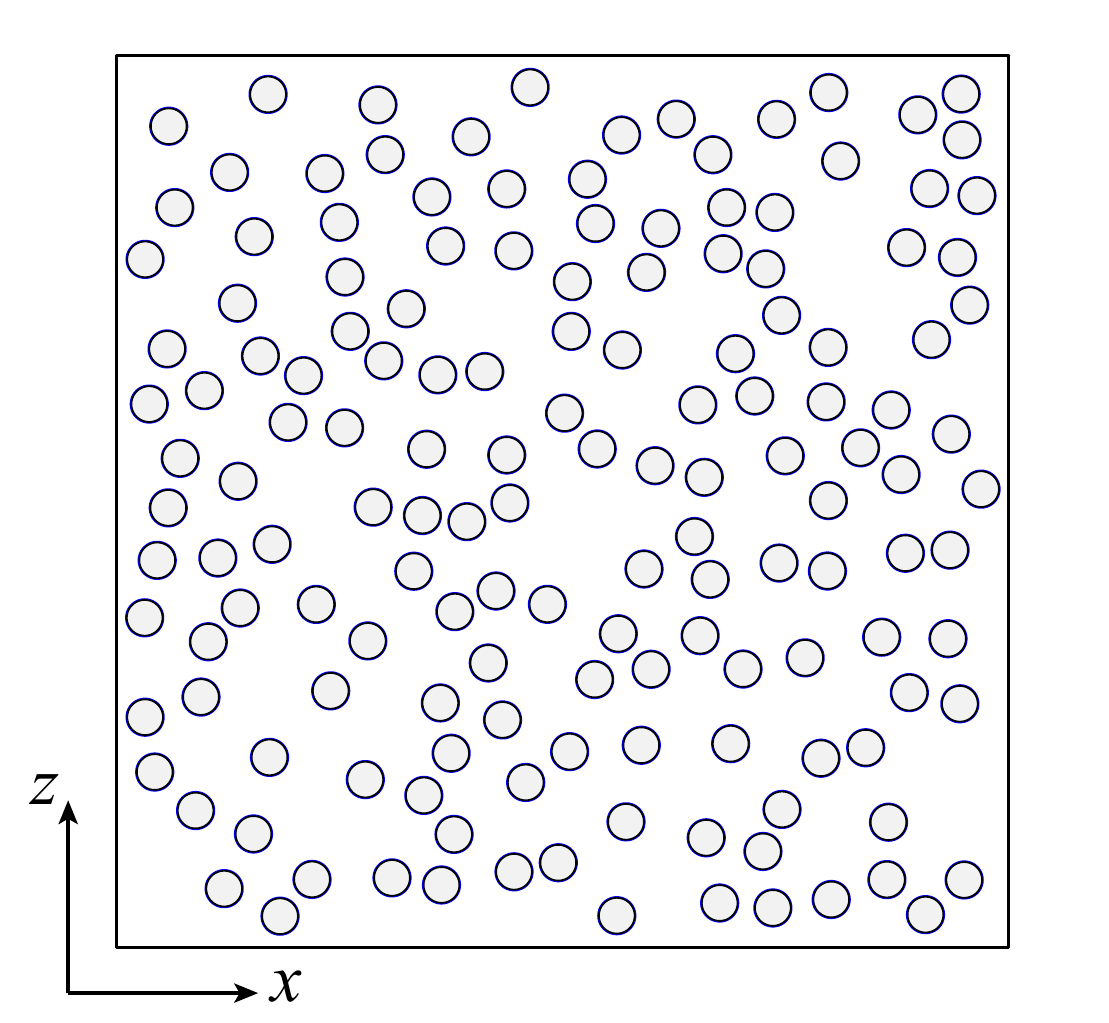}
	\caption{The schematic diagram of the 2D RB cell with porous media. In the cell, $N=150$ circular obstacles of diameter $D=0.04$ are placed randomly, with the condition that the minimum distance between any two obstacles satisfies $l\ge0.01$. The whole porosity of the system is $\phi=1-\frac{N\pi D^2}{4} =0.812$.}
	\label{fig:model}
\end{figure}

A two-dimensional square RB cell with length $L$ is considered in our simulation. The temperature difference between the hot lower plate and the cold upper plate is set to be $\Delta$. A set of randomly placed circular obstacles is used to model the porous media inside, as shown in \hyperref[fig:model]{figure~1}. For the boundary, no-slip and isothermal conditions are applied at the lower and upper plates, while no-slip and adiabatic conditions are applied at two side walls. Meanwhile, the surface of obstacles is regarded as no-slip and heat-conducting. The numerical methods applied to this system are shown as follows.

\subsection{Governing equations of the flow}

Based on the Oberbeck–Boussinesq approximation, the governing equations of the flow in the pore read:
\begin{equation}\label{Eqflow}
\begin{aligned}
\boldsymbol{\nabla}\cdot\boldsymbol{u}&=0,\\
\frac{\partial\boldsymbol{u}}{\partial t}+\boldsymbol{u}\cdot\boldsymbol{\nabla u}&=-\boldsymbol{\nabla} p+\sqrt{\frac{Pr}{Ra}}\nabla^2\boldsymbol{u}+T\boldsymbol{e_z}+\boldsymbol{f},\\
\frac{\partial T}{\partial t}+\boldsymbol{\nabla}\cdot(\boldsymbol{u}T)&=\sqrt{\frac{1}{Ra\cdot Pr}}\nabla^2 T,\\
\end{aligned}
\end{equation}
where $\boldsymbol{u}=(u,w)$ is the velocity vector, $p$ is the pressure field, $T$ is the temperature, $\boldsymbol{e_z}$ is the unit vector along the vertical direction, and $\boldsymbol{f}$ represents the resistance force to the moving fluid by fixed circular obstacles. Scaled quantities, including $L$ for length, $\Delta$ for temperature, $U=\sqrt{g\beta\Delta L}$ for velocity, and $L/U$ for time are used to non-dimensionalize the governing equation, where $g$ is the gravitational acceleration, and $\beta$ is the coefficient of thermal expansion of the fluid.

The Rayleigh number $Ra={g\beta\Delta L^3}/{(\nu\kappa_f)}$  and the Prandtl number $Pr={\nu}/{\kappa_f}$  are two non-dimensional parameters that control the system, where $\nu$ is the kinematic viscosity and $\kappa_f=\lambda_f/(\rho C_p)$ is the thermal diffusivity of the fluid. $\lambda_f,\rho,C_p$ are the thermal conductivity, density, and thermal capacity of the fluid, respectively. The response parameters include the Nusselt number $Nu=\sqrt{RaPr}\langle wT\rangle_{x,t}-\langle\partial_zT\rangle_{x,t}$, measuring the total heat transfer and the Reynolds number $Re= \sqrt{Ra/Pr} \sqrt{\langle |\boldsymbol{u}|^2 \rangle_{V,t}}$, measuring the strength of convection, where $\langle\cdot\rangle_{x,t}$ denotes the average over time and a horizontal plane, and $\langle\cdot\rangle_{V,t}$ denotes the average over time and space. 

\subsection{Numerical treatment of porous media}

To consider the effect of the porous media on the momentum equations, a direct-forcing immersed boundary method (IBM) is adopted, in the Euler–Lagrange framework \citep{uhlmann2005immersed,breugem2012second,wang_heat_2021}. In each step, a prediction velocity is obtained first by advancing the momentum equations without considering the IBM force $\boldsymbol{f}$. Then the first prediction velocity is interpolated from the Eulerian grid to a Lagrangian grid distributed uniformly along the boundary of circular obstacles. The IBM force $\boldsymbol{f}$ is computed for satisfying no-slip and no-penetration conditions on the Lagrangian grid and then spread back to the Eulerian grid using the moving-least-squares approach \citep{VANELLA2009AMoving,DETULLIO2016AMov}. Finally, the force is used to update the velocity and correct the pressure distribution.

Moreover, the thermal effect of porous media is realized by solving the temperature equation in both two phases \citep{ardekani2018heat,ardekani2018numerical,sardina2018buoyancy}. A phase indicator $\xi$ is introduced to quantify the solid volume fraction. A level-set function $\zeta$, given by the signed distance to the obstacle surface ($\zeta$ is negative inside the obstacle and positive outside the obstacle), is adopted to computed  $\xi$ at each point. With $\zeta$ of four corner nodes, the phase indicator $\xi$ is determined using the formula \citep{ardekani2018heat,kempe2012improved}:
\begin{equation}
    \xi=\frac{\displaystyle\sum\limits^{n=1}_{4}{-\zeta_n\mathcal{H}(-\zeta_n)}}{\displaystyle\sum\limits^{n=1}_{4}{|\zeta_n|}},
\end{equation}
where $\mathcal{H}$ is the Heaviside step function. By using $\xi$, the combined velocity can be expressed as: 
\begin{equation}
\boldsymbol{u_{cp}}=\xi\boldsymbol{u_p}+(1-\xi) \boldsymbol{u_f},
\end{equation}
where $\boldsymbol{u_p}$ and $\boldsymbol{u_f}$ represent the velocity in porous media and fluid, respectively, and there is $\boldsymbol{u_p}=\boldsymbol{0}$. Similarly, the combined thermal conductivity is expressed as:
\begin{equation}
\lambda_{cp}=\xi \lambda_p+(1-\xi) \lambda_f,
\end{equation}
where $\lambda_p$ and $\lambda_f$ represent the thermal conductivity in porous media and fluid, respectively. We assume $\rho C_p$ to be the same in the fluid and porous media for simplification. The temperature equation becomes:
\begin{equation}\label{Eqtemp}
    \frac{\partial T}{\partial t}+\boldsymbol{\nabla}\cdot(\boldsymbol{u_{cp}}T)=\sqrt{\frac{1}{Ra\cdot Pr}}\boldsymbol{\nabla}\cdot({\tilde{\lambda_{cp}}\boldsymbol{\nabla} T}),
\end{equation}
where we use the relative combined conductivity $\tilde{\lambda_{cp}}=\lambda_{cp}/\lambda_f$.

By coupling the IBM method and two-phase heat transfer, the dynamic and thermal effects of porous media on the flow and heat transfer are realized. More details of the coupling method we used on this system are shown in the previous work \citep{liu2020rayleigh}.

\subsection{Numerical details}
For analyzing the model numerically, we construct a uniform, staggered Cartesian grid, and discretize the governing equation in space by the second-order central finite-difference method. The time-stepping of the explicit terms is based on a fractional-step third-order Runge–Kutta scheme, and the implicit terms are based on a Crank–Nicolson scheme with a pressure correction step set following. For more details on the numerical schemes of the governing equations, we refer the reader to van der Poel {\it et al.} \citep{van2015pencil}.

In our simulation, $Pr=4.3$ is taken for water, and $Ra$ varies in the range of $[10^7,10^9]$. The ratio of thermal conductivity $\lambda=\lambda_p/\lambda_f$  varies from 0.1 to 50, representing the thermal conductivity from plastic to metal for porous media. The geometry holds the same as in \hyperref[fig:model]{figure~1} for all cases, and a uniform $1080\times1080$ grid is taken to achieve a full resolution of both the obstacles and the boundary layers, with each obstacle diameter resolved by 43 nodes and the thermal boundary layer described by at least 10 grid points. The non-dimensional maximum time step adopted ranges from $\Delta t=2\times10^{-4}$ to $1\times10^{-3}$, depending on the value of $\lambda$ and $Ra$. The simulations are run over at least 1000 non-dimensional time units after the system has reached the statistically stationary state to obtain good statistical convergence. The relative difference of $Nu$ based on the first and second halves of the simulations is generally less than $2\%$ \citep{stevens2010jfm}.

\section{Result and Discussion\label{sec3}}
In this section, to form an overall understanding of the thermal effect of porous media, we perform analysis from various aspects, including heat transfer, global statistics of temperature and velocity fields, plume behaviors, and the dominated thermal dissipation regime.

\begin{figure}
	\centering
	\includegraphics[width=0.8\linewidth]{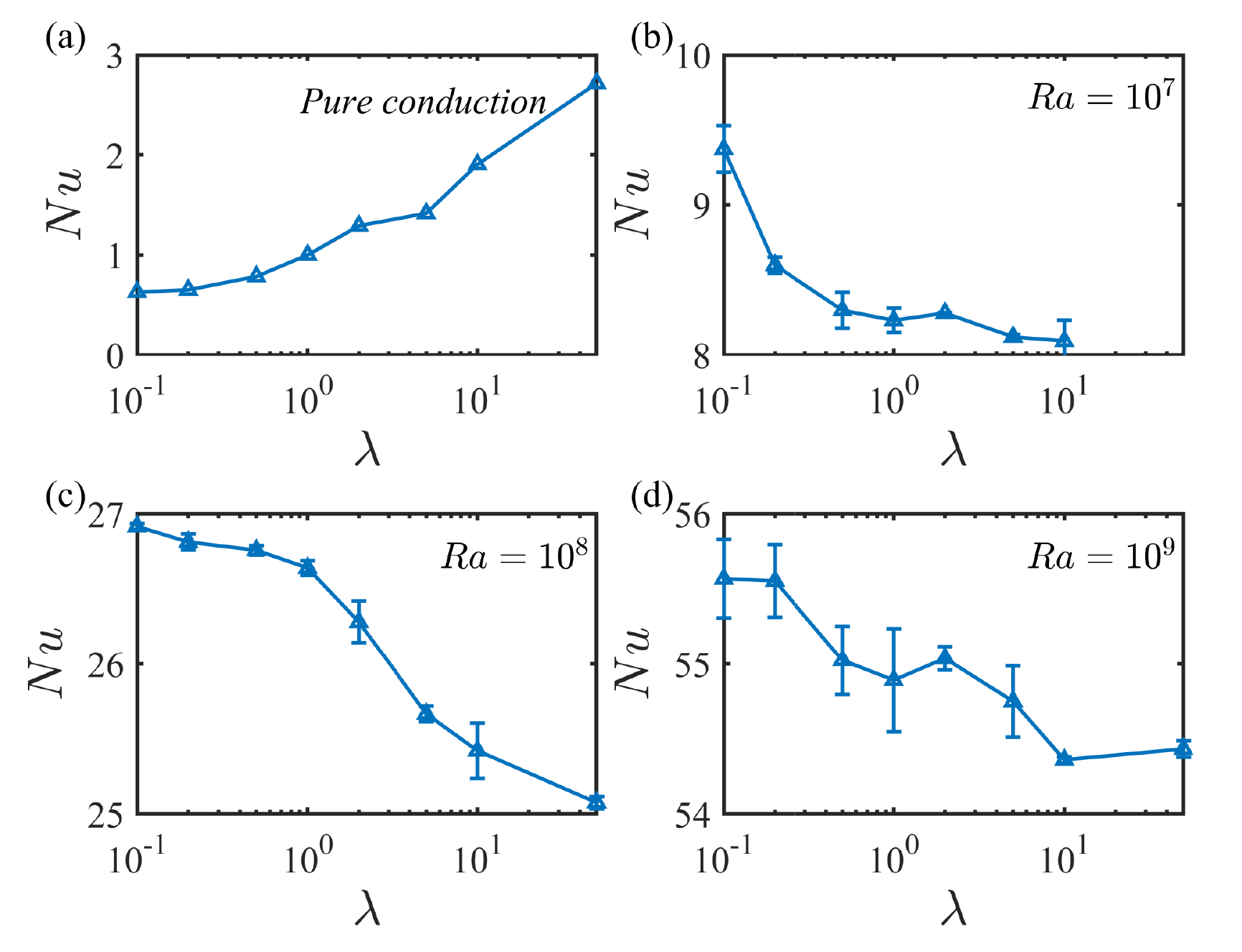}
	\caption{ (a) Variation of $Nu$ with $\lambda$ under pure conduction. (b-d) Variation of $Nu$ with $\lambda$ at different $Ra$.}
	\label{fig:Nuk}
\end{figure}

\begin{figure}
	\centering
	\includegraphics[width=0.5\linewidth]{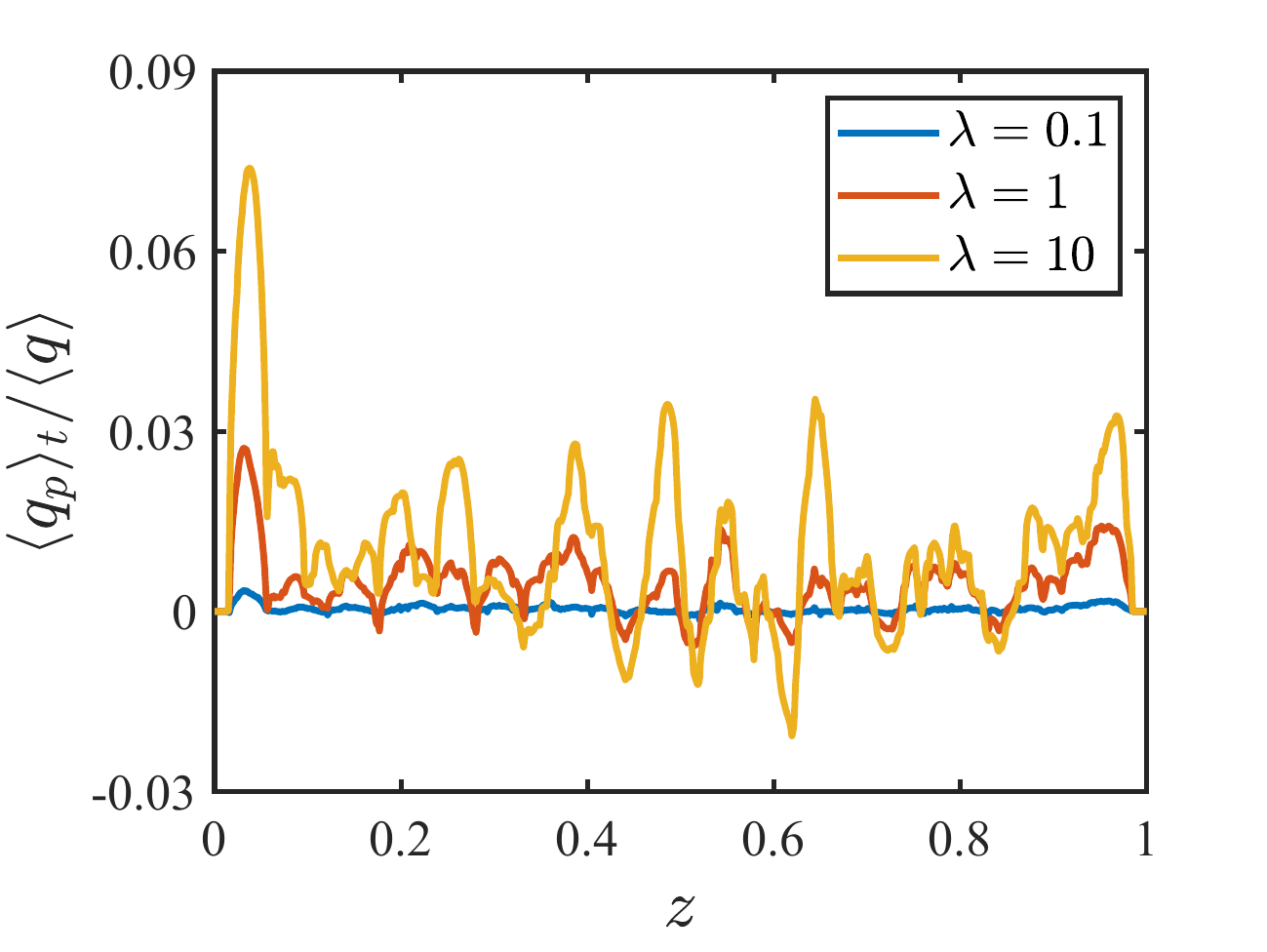}
	\caption{The ratio of time-averaged vertical heat flux through obstacles to the total heat flux at different horizontal planes, at $Ra=10^8$ and various thermal conductivity $\lambda=0.1,1,10$. }
	\label{fig:condheat}
\end{figure}

\subsection{Global statistics}

The variation of thermal conductivity of the circular  obstacles seems to have a small effect on heat transfer for the parameter considered, as shown in \hyperref[fig:Nuk]{figure~2}. The results of $\lambda=1$ cases are consistent with the results in the literature \citep{liu2020rayleigh}. In the pure conduction case (\hyperref[fig:Nuk]{figure~2}(a)), it is obvious that the heat transfer grows as $\lambda$ increases, but the growth is not large in contrast to the great variation of $\lambda$, due to the separation of the obstacles. However, when the fluid convection on participates, with $Ra$ from $10^7$ to $10^9$, the total heat transfer shows an opposite trend as shown in \hyperref[fig:Nuk]{figures~2}(b-d). One would expect to see $Nu$ enhancement with increasing $\lambda$, surprisingly here we observe an opposite trend. In all three convection cases, out of expectation, $Nu$ decreases with $\lambda$, though the change is relatively small compared to the total heat transfer. For example, as illustrated in \hyperref[fig:Nuk]{figure~2}(c), at $Ra=10^8$, with the thermal conductivity $\lambda$ growing from $0.1$ to $50$, $Nu$ decreases from $26.92$ to $25.08$, given a relative decrease of $6.84\%$. Furthermore, as the convection becomes stronger, the relative decrease of $Nu$ is reduced to $2.04\%$ at $Ra=10^9$. In order to understand this unexpected trend, more detailed information on the flow structure and temperature field is studied in the following parts.

The total heat transfer in the system can be divided into convection and conduction, and first we consider the heat conduction effect of porous media. For simplification, we pick typical cases $\lambda=0.1,1,10$ under $Ra=10^8$  in the analysis after. In order to quantify the contribution of porous media to thermal conduction, we calculate the ratio of time-averaged heat flux through obstacles to total heat flux at different horizontal planes, expressed by the non-dimensional quantities:
\begin{equation}
\langle q_p\rangle_t/\langle q\rangle_t=\langle \int{\lambda\frac{\partial T}{\partial z}dx_p}\rangle_t/Nu,
\label{flux-ratio}
\end{equation}
where $dx_p$ denotes the dimensionless horizontal differential step in the porous media. The results are presented in \hyperref[fig:condheat]{figure~3}. As the cross section area of porous media is highly dependent on the height of horizontal section in our system, the relative heat flux through obstacles $\langle q_p\rangle_t/\langle q\rangle_t$ varies a lot with $z$. There are two peaks near the top and bottom boundaries, caused by the great temperature gradient in two boundary layers. Furthermore, by comparing the cases of different thermal conductivity, we note that the increase in thermal conductivity $\lambda$ brings a relatively high increase in the conduction heat flux of obstacles, but the contribution to total heat transfer is still small even for $\lambda=10$. As expected, the heat transfer is convection-dominated, and the enhancement due to heat conduction is minor.

\begin{figure}
	\centering
	\includegraphics[width=0.9\linewidth]{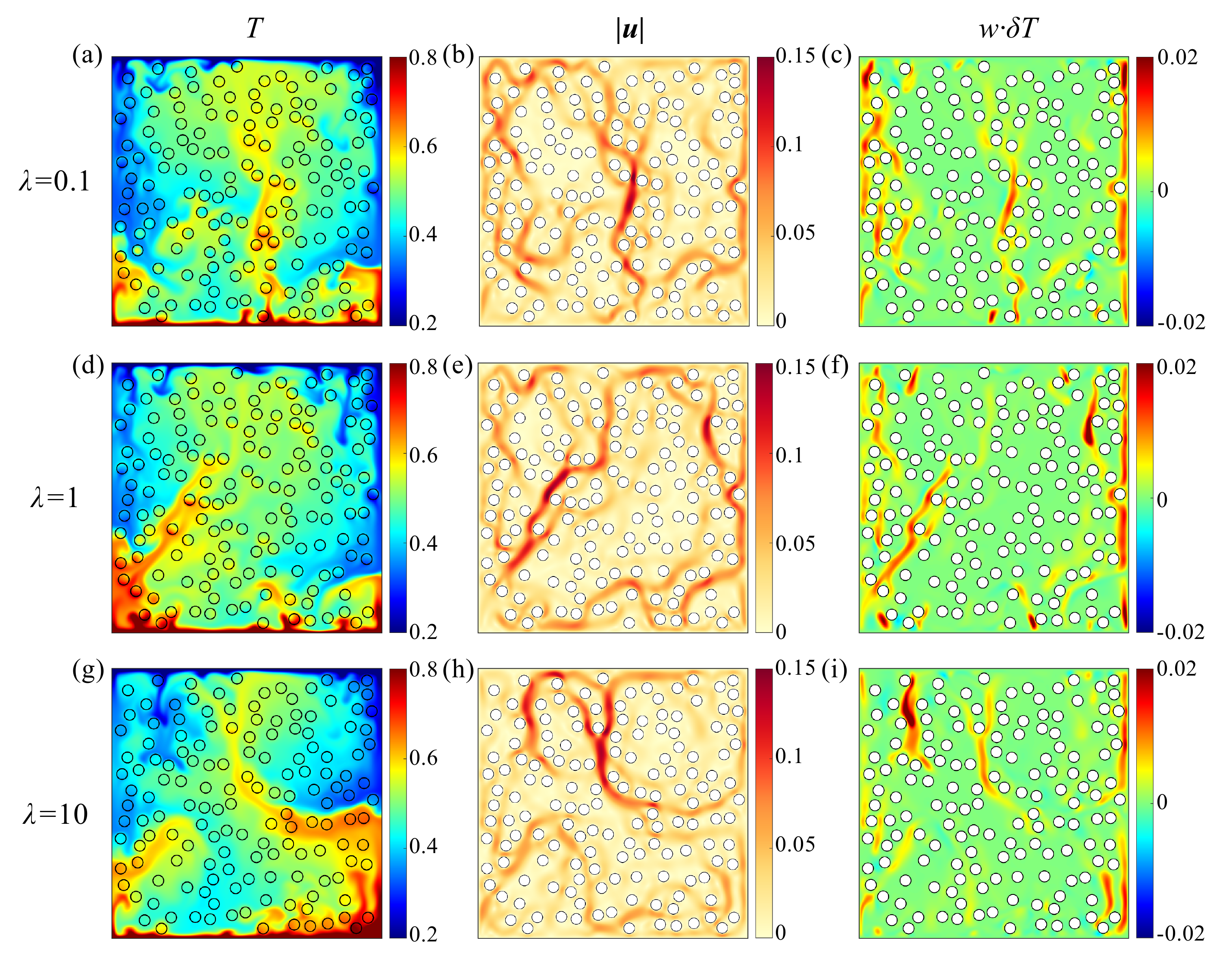}
	\caption{Typical snapshots of the instantaneous temperature $T$ (a,d,g), the velocity magnitude $|\boldsymbol{u}|$ (b,e,h) and the local convective heat flux $w\cdot \delta T$ (c,f,i) (where $\delta T=T-0.5$) at $Ra=10^8$, $\lambda=0.1,1,10$. Circles in the figures indicate the randomly displayed obstacles.}
	\label{fig:temp}
\end{figure}

For convective heat transfer, the statistics of the temperature and velocity fields can give essential information. The typical snapshots of the instantaneous temperature fields, the velocity magnitude fields and the local convective heat flux fields are displayed in \hyperref[fig:temp]{figures~4}(a-i). Because of the obstruction of randomly distributed obstacles, the LSC is suppressed, and the regular convection channels in Ref. \citep{liu2020rayleigh} are not formed as well. Thermal plumes are detached from the boundary layers, fragmented by the obstacles, and transport heat through several tortuous and discontinuous channels, as shown in the heat flux field \hyperref[fig:temp]{figures~4}(c,f,i). As thermal conductivity varies, on the whole, there are not many differences between cases of different $\lambda$. Under lower $\lambda$, more thermal residual in the obstacles can be seen at the point of contact with the plumes, where the temperature is higher than the surrounding temperature. Moreover, as the obstacles of low thermal conductivity prevent heat from passing through them, the temperature field seems a bit more fragmentized and plumes take up less area. The relation between plume area and thermal conductivity will be discussed later, in section \ref{pls}. Moreover, although the distribution of obstacles is the same, the flow channels are quite different between the cases of different $\lambda$, as shown in the velocity magnitude fields \hyperref[fig:temp]{figures~4}(b,e,h). There are more strong heat flux channels at $\lambda=0.1$, contributing to a stronger total heat transfer; while at $\lambda=10$, the flow strength seems weaker, which will be confirmed later by comparing the Reynolds number; the weaker flow depresses the heat transfer through flow channels surrounded by obstacles.

\begin{figure}
	\centering
	\includegraphics[width=0.5\linewidth]{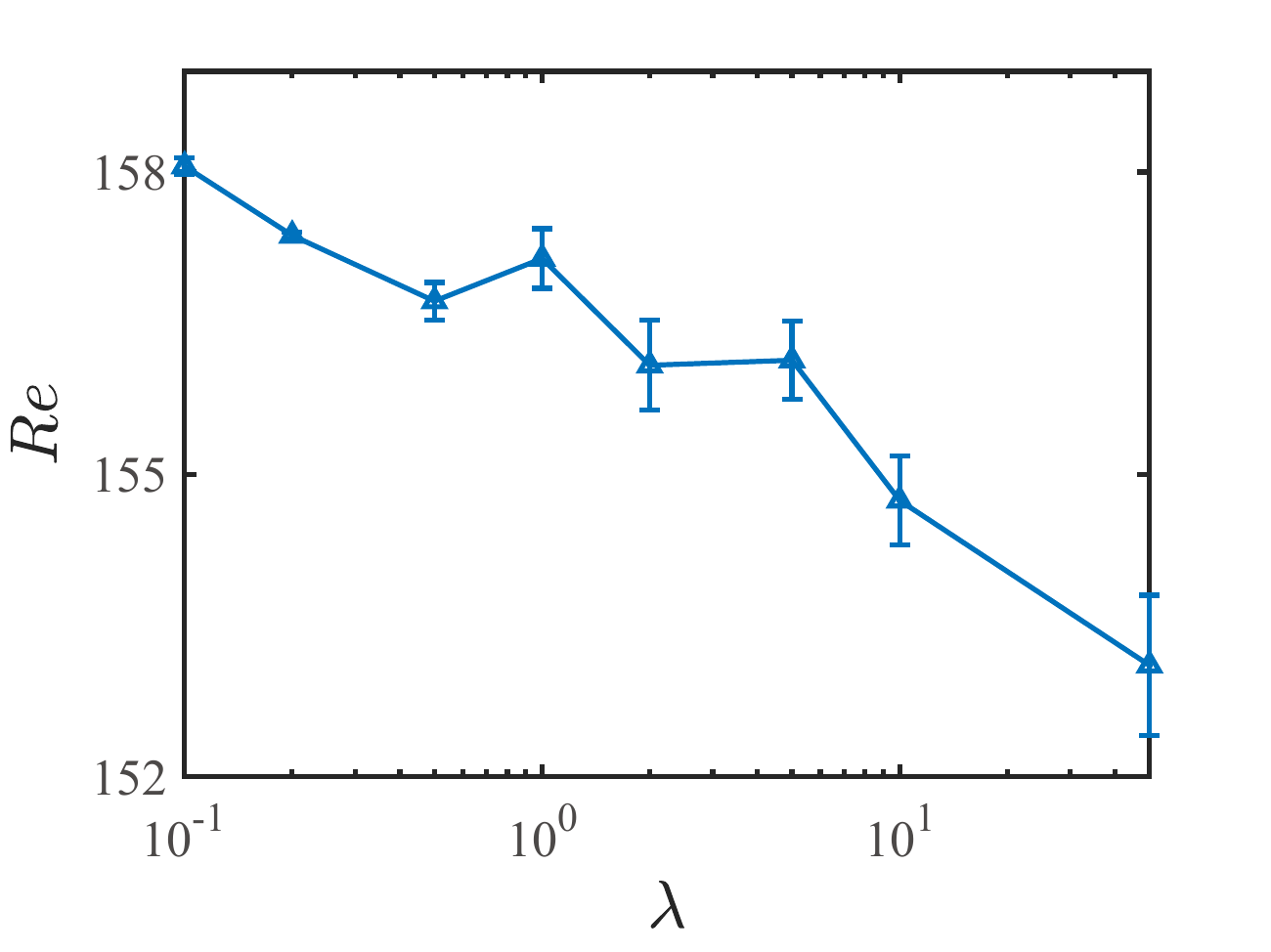}
	\caption{Variation of $Re$ with $\lambda$, at $Ra=10^8$.}
	\label{fig:Rek}
\end{figure}

An important quantity describing the convection strength of the flow field is the Reynolds number, $Re$. As illustrated in \hyperref[fig:Rek]{figure~5}, the Reynolds number tends to be slightly reduced as the porous media become more heat-conductive, which means the convection is suppressed to some extent. With $\lambda$ increasing from $0.1$ to $50$, $Re$ is reduced about $3.14\%$. The reduction of convection strength may have an effect on total heat transfer through plumes, which are the main heat carrier in RB convection. 
\subsection{Plume behaviors\label{pls}}

\begin{figure}
	\centering
	\includegraphics[width=1.0\linewidth]{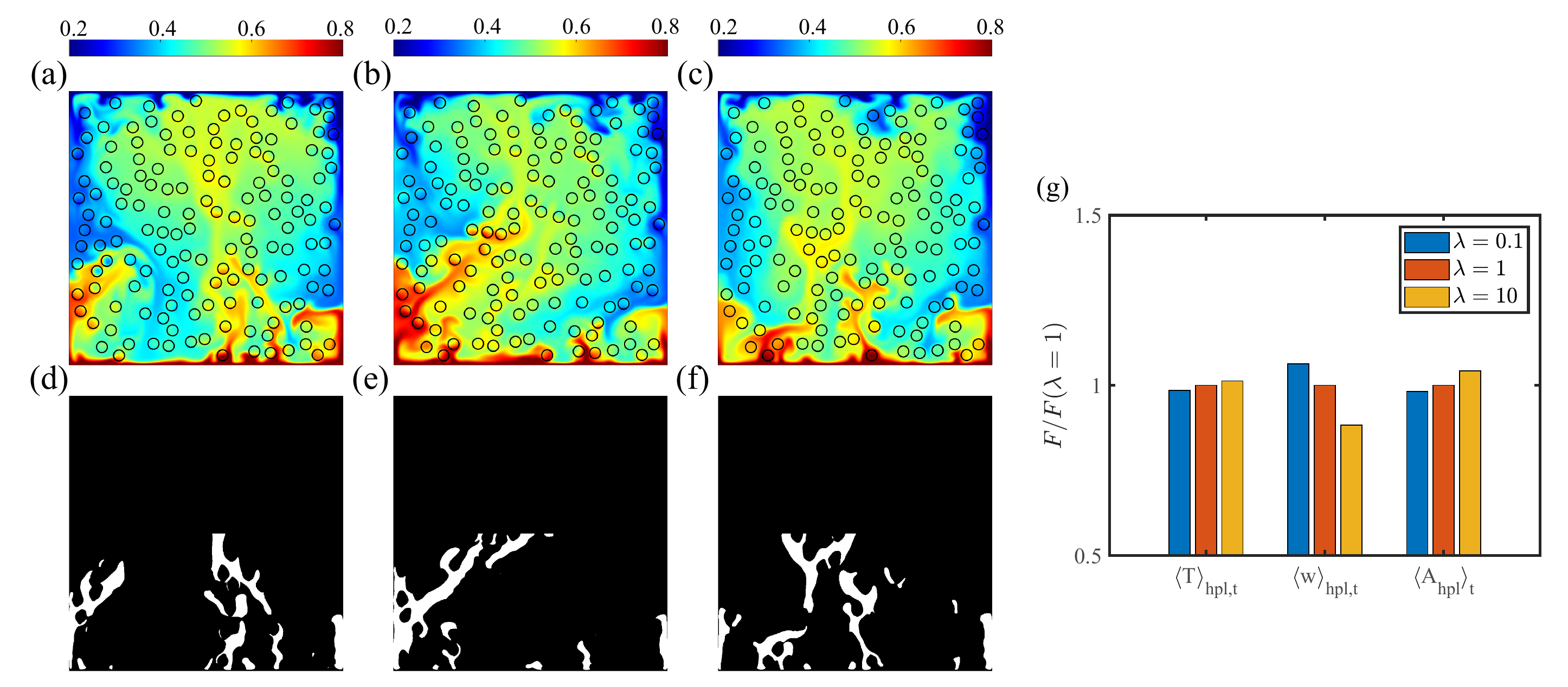}
	\caption{(a-c) Snapshots of the instantaneous temperature field at $Ra=10^8, \lambda=0.1,1,10$. Thermal plumes can be observed clearly. (d-f) The corresponding hot plumes recognized for the moment of snapshots (a-c), following the rules given in van der Poel \textit{et al}. \citep{Vanderpoel2015Plume}. (g) The average temperature $\langle T\rangle_{hpl,t}$, the average vertical velocity $\langle w\rangle_{hpl,t}$, and the average area $\langle A_{hpl}\rangle_{t}$ of hot plumes for $Ra=10^8$ and $\lambda=0.1,1,10$. The results are normalized using the data in case $\lambda=1$, as $\langle T\rangle_{hpl,t}(\lambda=1)=0.6443$, $\langle w\rangle_{hpl,t}(\lambda=1)=0.0424$ and $\langle A_{hpl}\rangle_{t}(\lambda=1)=0.0554$.}
	\label{fig:hotplume}
\end{figure}

It is known that thermal plumes are the main heat carrier of heat transfer in the RB system. To identify thermal plumes from the turbulent background, we use the algorithm proposed by van der Poel {\it et al.} \citep{Vanderpoel2015Plume}: $\pm(\theta(x,z)-\langle \theta(x,z)\rangle _x)>c\cdot \theta_{rms}$ (where $+$ is for hot plumes and $-$ for cold plumes); $\sqrt{Ra\cdot Pr}\cdot w(x,z)\cdot \theta(x,z)>c\cdot Nu$. In the algorithm, $\theta(x,z)=T(x,z)-0.5$ as we use $T\in[0,1]$ while the calculation of vertical convective heat flux requires $\theta\in[-0.5,0.5]$, and $\theta_{rms}$ is the root-mean-square temperature over the horizontal section. In the current situations, we choose the empirical parameter $c=0.8$, which is the same as one used by Huang {\it et al.} \citep{Huang2013confine}. Moreover, Only hot plumes in the lower half of the domain and cold plumes in the upper half are considered \citep{Vanderpoel2015Plume,Jiang2018Control}. Some typical results of hot plume detection are shown in \hyperref[fig:hotplume]{figures~6}(a-f). It is clear that hot plumes are detached from the boundary layer, flow upward and meet the obstacles. The extracted hot plumes in \hyperref[fig:hotplume]{figures~6}(d-f) are highly consistent with the high temperature region in \hyperref[fig:hotplume]{figures~6}(a-c), and the obstacles surrounded by plumes are well recognized, indicating this plume extraction algorithm works well. Next, at $Ra=10^8$, the average temperature $\langle T\rangle_{hpl,t}$, the average vertical velocity $\langle w\rangle_{hpl,t}$, and the average area $\langle A_{hpl}\rangle_{t}$ of hot plumes are counted, as displayed in \hyperref[fig:hotplume]{figure~6}(g). As $\lambda$ increases, the plume temperature and plume area increase as well, while the plume velocity is decreased a lot, about $16.94\%$ from $\lambda=0.1$ to $\lambda=10$, much larger than the relavite reduction of $Nu$. The increase in the plume area may be caused by the smoothing effect of porous media, as the porous media allows the heat of the high temperature fluid to be carried to the other end, making the high-temperature region larger. The reduction of the plume velocity impedes total heat transfer, while the expansion of the plume area promotes heat transfer. These two effects counteract each other to some extent, weakening the overall effect of $\lambda$ on total heat transfer efficiency.

During the lifetime of a hot plume, it is detached from the boundary layer, moves across the bulk region, interacts with the circular obstacles, and finally reaches the top plate. In this process, the heat transfer is influenced by the porous media directly through the interaction between obstacles and plumes. Consequently, as the dynamic interaction is well studied by previous studies \citep{liu2020rayleigh,liu2021lagrangian,ataei2019flow}, here we focus on the heat exchange between porous media and fluid, which is important in the analysis of the thermal effect. The mutual effect of plumes and porous media is first influenced by the obstacles the plume passes through. Due to the no-slip boundary conditions, the flow decelerates near the the surface of obstacles, ruled not to be a part of plumes. Therefore, we consider a region near each obstacle, and define a contact parameter $s$ as the ratio of the plumes' area in this region to the total area. This region is selected as the distance to the obstacle's center $0.55D\le d\le 0.6D$, where the maximum contact parameter (when an obstacles is encircled by a plume) reaches $s_{max}=91.4\%$. 

\begin{figure}
	\centering
	\includegraphics[width=0.9\linewidth]{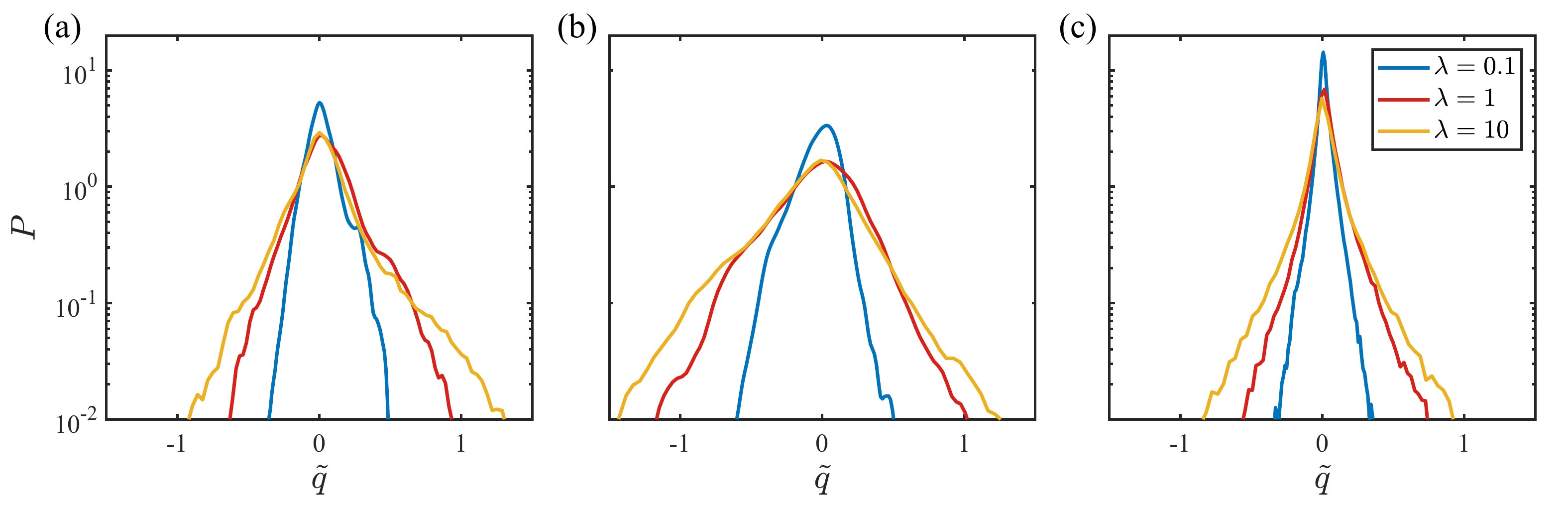}
	\caption{PDFs of dimensionless heat exchange rate of the obstacle ($\tilde{q}=\oint \tilde{\lambda_{cp}}\boldsymbol{\nabla}T\cdot\boldsymbol{n}dl$) with fluid when (a) cold plumes pass the obstacle (b) hot plumes pass the obstacle (c) no plume passes the obstacle. The blue, red, yellow lines refer to the case $\lambda=0.1, 1, 10$, respectively. Positive $\tilde{q}$ means heat release from obstacles to the fluid. $Ra=10^8$.}
	\label{fig:heatexchange}
\end{figure}

For every moment, we compute $s$ to each obstacle, and define that the obstacle is passed by one kind of plumes once $s>10\%$. Meanwhile, one obstacle is considered to have no plume flowing through it at one moment when $s$ for hot plumes and cold plumes are both absolutely $0$. The total heat exchange rate of each obstacle with the fluid is also calculated by a curvilinear integral of heat flux over its surface, $\tilde{q}=\oint \tilde{\lambda_{cp}}\boldsymbol{\nabla}T\cdot\boldsymbol{n}dl$ in dimensionless form, where $dl$ denotes the dimensionless horizontal differential step on the surface of the obstacles and $\boldsymbol{n}$ is the unit vector normal to the surface. Larger $\tilde{q}$ means that the obstacles exchange heat more, and may have a greater influence on the heat transfer. Data with 150 obstacles in our model, and over 600 snapshots covering a period $\Delta t>300$ for each case are processed, and the probability density functions (PDF) are illustrated in \hyperref[fig:heatexchange]{figure~7}. When no plume passes the obstacles, the PDFs are symmetrical as the obstacles exchange heat with the random turbulent background. When cold plumes pass the obstacles, PDFs shifts to the right side, meaning that the obstacles tend to release heat to cold plumes; similarly, the obstacles tend to absorb heat from hot plumes as PDFs shifts to the left side. In all situations, whether or not the plume passes through the obstacles, the PDF curves of large thermal conductivity is wider, showing a larger variance of $\tilde{q}$, which can be observed quantitatively by the standard deviation of $\tilde{q}$ in the \hyperref[tab1]{table~1}. The standard deviation of $\tilde{q}$ increases significantly as $\lambda$ increases. Moreover, when plumes pass the obstacles, compared to the case of no plume, the heat exchange between the obstacle and the fluid becomes more intense. As $\lambda$ increases, the total amount of heat exchange is raised more when there is a plume. The heat exchange between the hot and cold plumes passing through the obstacles is not the same, due to the asymmetry of the obstacles distribution. The results reveal that the passage of the plume increases the heat exchange between the obstacles and the fluid to a great extent; the increase of thermal conductivity $\lambda$ can promote the heat exchange when plumes pass, as the standard deviation of $\tilde{q}$ is significantly increased.

To further investigate the total heat exchange between obstacles and fluid, we analyze how often and how large area the plume makes contact with the obstacles. In our count, the situation of no plume passing ($s=0$) takes $74.39\%, 73.05\%, 73.84\%$ of all data for $\lambda=0.1, 1, 10$, respectively, and the situation of hot plumes or cold plumes ($s>10\%$) passing takes $21.23\%, 22.23\%, 21.45\%$. The non-dimensional total mean contact surface  $\langle s\rangle_{N,t}=9.04\%, 9.19\%, 8.81\%$ for three cases. The data shows that plumes can only interact with few obstacles, and the contact surface area is also limited. The increase of $\lambda$ enhances the heat exchange when a plume passes an obstacles observably, but the contacts between the plume and the obstacles remain at a low level. Therefore, although the increase of thermal conductivity can promote heat exchange with plumes, the low frequency of plume-obstacle interaction limits the effect on total heat transfer.

\begin{table}
	\centering
	\def~{\hphantom{0}}
	\begin{tabular}{|c|ccc|}
	\hline        
		Cases & $\lambda=0.1$ & $\lambda=1$ & $\lambda=10$ \\ \hline
		cold plumes & $0.11$ & $0.21$ & $0.26$ \\ 
		hot plumes	& $0.14$ & $0.30$ & $0.36$\\ 
		no plume    & $0.06$ & $0.12$ & $0.17$\\ \hline
	\end{tabular}%
\bigskip
	\caption{The standard deviation of the obstacle's heat exchange rate $\tilde{q}$ when cold plumes/hot plumes/no plume pass at $\lambda=0.1,1,10$. $Ra=10^8$. These results correspond to the PDFs in figure \ref{fig:heatexchange}.}
    \label{tab1}
\end{table}

\begin{figure}
	\centering
	\includegraphics[width=0.8\linewidth]{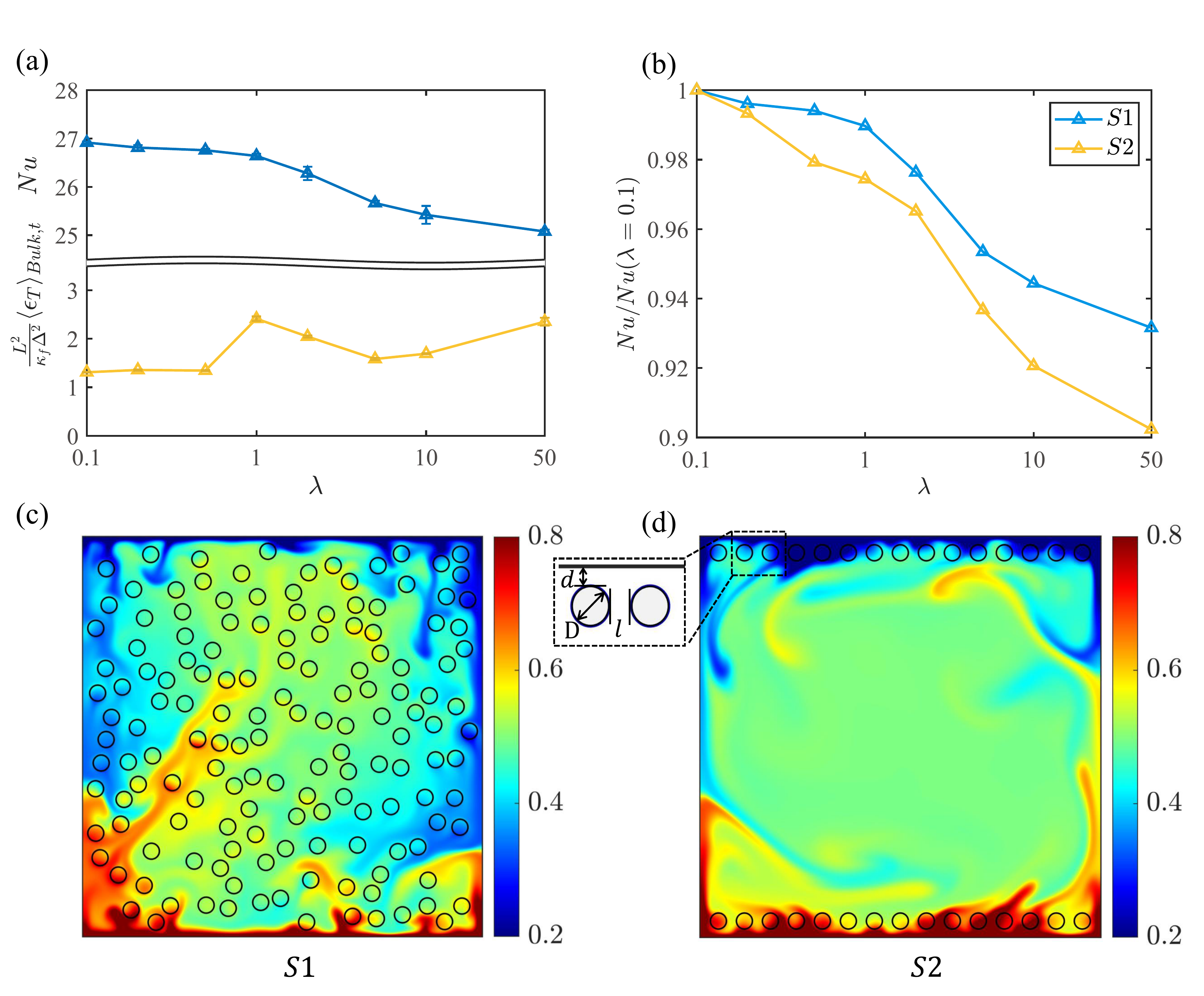}
	\caption{(a) Variation of dimensionless time-and-space-averaged thermal dissipation  in the bulk region $\frac{L^2}{\kappa_f\Delta^2}\langle\epsilon_T \rangle_{Bulk,t}$ with $\lambda$ (the yellow line), compared to the Nusselt number (the blue line). $Ra=10^8$.(b) Variation of $Nu$ with $\lambda$ at $Ra=10^8$ in the old system ($S1$, the blue line, $\phi=0.812$) and the new system ($S2$, the yellow line, $\phi=0.962$), both normalized by the value at $\lambda=0.1$. For $S1$, $Nu(\lambda=0.1)=26.92$; for $S2$, $Nu(\lambda=0.1)=20.17$. (c) The instantaneous temperature field of the old system $S1$, at $Ra=10^8$ and $\lambda=1$. (d) The instantaneous temperature field of the new system $S2$, at $Ra=10^8$ and $\lambda=1$. Two rows of $15$ circular obstacles of diameter $D=0.04$ are placed symmetrically and horizontally on the upper and lower sides of the RB cell, $d=0.02$ from the boundary, with equal distances $l=0.025$ between obstacles. The whole porosity of the system is $\phi=0.962$.}
	\label{fig:NuBL}
\end{figure}
\subsection{Dominated regime}

Finally, let us turn to thermal dissipation rate distribution. In the Rayleigh–Bénard convection, the overall thermal dissipation is related to the Nusselt number by exact relation: $\langle\epsilon_T \rangle_{V,t}=(\kappa_f\Delta^2/L^2)Nu$, and the thermal dissipation rate is defined as $\epsilon_T=(\kappa_{cp}\Delta^2/L^2)|\nabla T|^2$, where $\kappa_f=\lambda_f/\rho C_p$ and $\kappa_{cp}=\lambda_{cp}/\rho C_p$. The exact relation of the two-phase RB flow in our system is derived in appendix \ref{appA}. Choosing the bulk region as $0.3<z<0.7$, we calculate the average thermal dissipation rates over time and space in the bulk region of the system, and the results are shown in \hyperref[fig:NuBL]{figure~8}(a). According to the exact relation, the sum of the averaged thermal dissipation in the bulk region and in the thermal boundary layer is the Nusselt number. The figure clearly shows that the bulk region contributes a small part to Nu, which means the heat transfer is dominated by boundary layers under the present parameter range \citep{grossmann_scaling_2000,grossmann_thermal_2001}. 


Actually, in our system, most obstacles are placed in the bulk region rather than in the boundary layer for all the cases, as the thickness of the boundary layer estimated by $\delta_{th}=1/(2Nu)$ is thin. Therefore, in a boundary-layer-dominated system, less distribution of obstacles in boundary layers may be one reason for the weak effect of $\lambda$ on $Nu$. To verify this conjecture, we remove the obstacles in the bulk region, and place all of them near the boundary layers regularly. For convenience, the old system and the new system are marked as $S1$ and $S2$, and the corresponding instantaneous temperature fields simulated at $Ra=10^8, \lambda=1$ are shown in \hyperref[fig:NuBL]{figures~8}(c,d), respectively. Two rows of $15$ obstacles are placed symmetrically and horizontally on the upper and lower sides of the RB cell, $d=0.02$ from the boundary, while the thickness of boundary layers $\delta_{th}=1/(2Nu)$ is in the range of $0.025-0.027$, calculated by the Nusselt number from the later simulations. Therefore, some part of obstacles is immersed into the boundary layers. The distances between obstacles are equal, as $l=0.025$. \hyperref[fig:NuBL]{Figure~8}(d) clearly shows that a LSC is formed, and shearing the inward-facing side of the obstacles. In the new system, most obstacles are indeed immersed into the boundary layers. As the thickness of the boundary layer is very thin, it is hard to place more obstacles in the boundary layers. 

The schematic diagram is investigated at $Ra=10^8, Pr=4.3$, and $\lambda\in[0.1, 50]$, and the heat transport curve is illustrated in \hyperref[fig:NuBL]{figure~8}(b), compared with the $Nu$ vs. $\lambda$ curve of the original system $S1$ under the same parameters. In both situations, though Nusselt number is found to decrease with increasing $\lambda$, the quantitative dependence is very different. When the porous media are placed in the boundary layer, the heat transfer has strong dependence with the thermal conductivity of the porous material. The significant drop in $Nu$ with $\lambda$ may result from two aspects: the obstacles in the boundary layers bring more resistance to the plumes’ detachment and motion, and the plume strength is reduced due to the reduction for temperature gradient in the boundary layer. As the number of obstacles in $S2$ is one-fifth of the number in $S1$, it is suggested that the obstacles in the boundary layer are much more efficient in the modification of the global heat transfer. 

\section{Conclusion\label{sec4}}

Based on the results and analysis above, the thermal effect of high-porosity porous media in our simulation set can be concluded. On the whole, the increase of thermal conductivity of circular obstacles suppresses convection strength and reduces total heat transfer. In terms of conduction, the vertical conduction heat transfer of obstacles is remarkably strengthened, but due to the discontinuity of obstacles and convection domination, this effect only weakly contributes to enhancing global heat transfer. For convection, plumes exchange more heat with obstacles when passing, which results in larger plume area, and lower plume velocity. However, the heat exchange between the thermal plumes and obstacles is limited due to the low frequency that plumes pass obstacles. Meanwhile, larger plume areas and lower plume velocities counteract each other to some extent. All of these restricted and opposite effects give a weakly decreasing trend of $Nu$ with $\lambda$. On the other hand, the thermal effect of the obstacles can be enhanced by putting them into the boundary layer. 

Constrained by the two-dimensional simulation model, the porous media can not be set as a continuous whole, but it can be realized in three-dimensional conditions. An important extension study in the future is extending the 2D model to 3D, constructing an interconnected porous media scheme to study the thermal effect on convection heat transfer. In addition, the region of the parameter can be extended as well, which may bring various interesting phenomena.
\clearpage
\begin{appendices}

\section{Derivation of the exact relation}\label{appA}
Considering the temperature equation with combined velocity and thermal conductivity, in non-dimensional form:
\begin{equation}\label{Eqapp}
    \frac{\partial T}{\partial t}+\boldsymbol{\nabla} \cdot(\boldsymbol{u_{cp}}T)=\sqrt{\frac{1}{RaPr}}\boldsymbol{\nabla}\cdot({\tilde{\lambda_{cp}}\boldsymbol{\nabla} T}),
\end{equation}
where $\tilde{\lambda_{cp}}=\lambda_{cp}/\lambda_f$ is the combined thermal diffusivity. Taking the product of Eq. \ref{Eqapp} with $T$, then averaging over the whole cell and a long time, one can obtain this equation:
\begin{equation}\label{Eqapp2}
    \frac{1}{2}\frac{d}{dt}\langle T^2\rangle_{V,t}+\frac{1}{2}\langle\boldsymbol{\nabla}\cdot(\boldsymbol{u_{cp}}T^2)\rangle_{V,t}=\sqrt{\frac{1}{RaPr}}\langle T\boldsymbol{\nabla}\cdot({\tilde{\lambda_{cp}}\boldsymbol{\nabla} T})\rangle_{V,t}.
\end{equation}
When the system reaches the the statistically stationary state, the first term on the left hand side equals to zero. Using the no-slip condition on the system boundary $\boldsymbol{u_{cp}}=\boldsymbol{u_f}=0$, the second term on the left hand side becomes:
\begin{equation}\label{Eqapp3}
    \frac{1}{2}\langle\boldsymbol{\nabla}\cdot(\boldsymbol{u_{cp}}T^2)\rangle_{V,t}=\frac{1}{2L^2}\langle\displaystyle\oint_{boundary}(\boldsymbol{u_{cp}}T^2\cdot\boldsymbol{n})\rangle_t=0,
\end{equation}
where $\boldsymbol{n}$ means the normal directions of the system boundaries. Meanwhile, the term in the right hand side of Eq. \ref{Eqapp2} can be expressed as:
\begin{equation}\label{Eqapp4}
    \langle T\boldsymbol{\nabla}\cdot({\tilde{\lambda_{cp}}\boldsymbol{\nabla} T})\rangle_{V,t}=\langle\boldsymbol{\nabla}\cdot(\tilde{\lambda_{cp}}T\boldsymbol{\nabla}T)\rangle_{V,t}-\langle\tilde{\lambda_{cp}}|\boldsymbol{\nabla}T|^2\rangle_{V,t}.
\end{equation}
Therefore, taking the Eq. \ref{Eqapp3} and Eq. \ref{Eqapp4} back, then the Eq. \ref{Eqapp2} can be written as:
\begin{equation}
\langle\boldsymbol{\nabla}\cdot(\tilde{\lambda_{cp}}T\boldsymbol{\nabla}T)\rangle_{V,t}=\langle\tilde{\lambda_{cp}}|\boldsymbol{\nabla}T|^2\rangle_{V,t}.
\end{equation}
As we define the thermal dissipation rate based on non-dimensional quantities, $\epsilon_T=(\kappa_{cp}\Delta^2/L^2)|\boldsymbol{\nabla}T|^2$, where $\kappa_{cp}=\tilde{\lambda_{cp}}\kappa_f$, the time and space average of it equals to a flux across the system boundary:
\begin{equation}        
\langle\epsilon_T\rangle_{V,t}=\frac{\kappa_f\Delta^2}{L^2}\langle\boldsymbol{\nabla}\cdot(\tilde{\lambda_{cp}}T\boldsymbol{\nabla}T)\rangle_{V,t}=\frac{\kappa_f\Delta^2}{L^2}\langle\displaystyle\oint_{boundary}(\tilde{\lambda_{cp}}T\boldsymbol{\nabla}T\cdot\boldsymbol{n})\rangle_t;
\end{equation}
on the boundary, $\kappa_{cp}=\kappa_f$ and $\tilde{\lambda_{cp}}=1$. With adiabatic condition on the side walls and isothermal condition on the hot lower plate and the cold upper plate, one can obtain the following relation:
\begin{equation}
\langle\epsilon_T\rangle_{V,t}=-\frac{\kappa_f\Delta^2}{L^2}\langle\partial_zT\rangle_{x,t,z=0}=\frac{\kappa_f\Delta^2}{L^2}Nu.
\end{equation}
Thus, the exact relation of the two-phase RB system is very similar to the exact relation of the single phase RB system.

\section{The robustness of the random distributions of obstacles}
\begin{figure}
    \centering
    \includegraphics[width=1.0\linewidth]{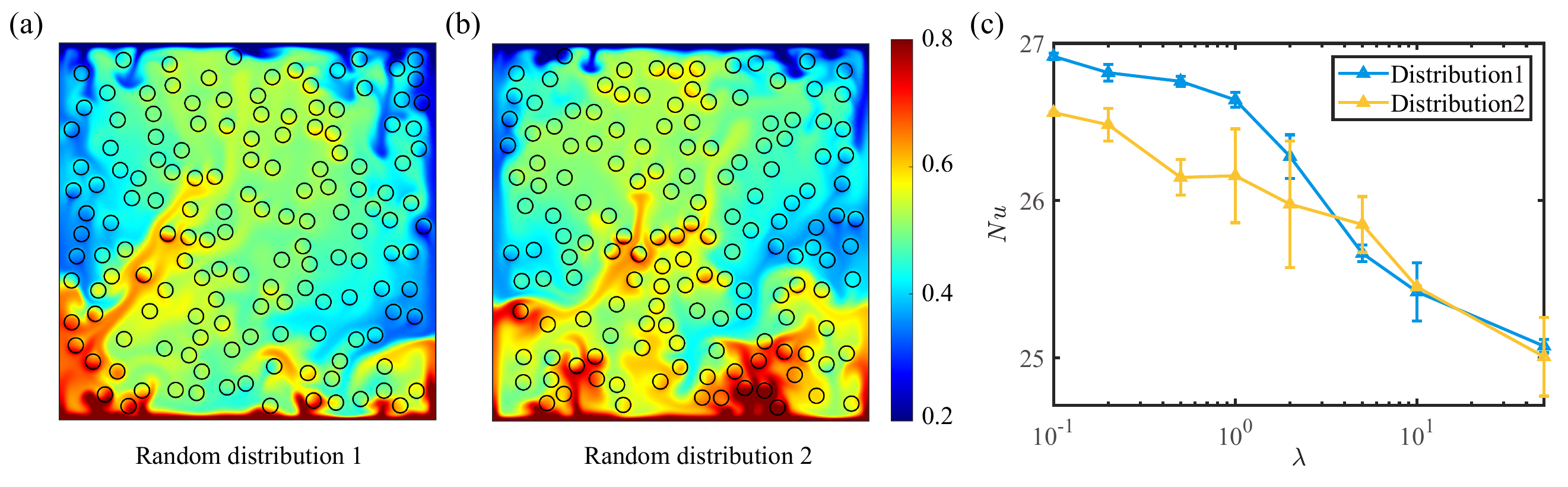}
    \caption{Two random distributions of obstacles and the corresponding heat transfer curves $Nu~\lambda$ under $Ra=10^8$. (a) The instantaneous temperature field of the Random distribution 1 (the old system discussed above) at $\lambda=1$,$Ra=10^8$. (b) The instantaneous temperature field of the Random distribution 2 (a new distribution) at $\lambda=1$,$Ra=10^8$. (c) Variation of $Nu$ with $\lambda$ at $Ra=10^8$ in the two systems with different random distributions of obstacles.}
    \label{fig:Randomcheck}
\end{figure}

Here, we will show the robustness of the random distributions of obstacles. Following the same random rules (the obstacle diameter $D=0.04$, the number of obstacles $N=150$, and the minimum distance between any two obstacles $l\ge0.01$), another random distribution of obstacles is generated. \hyperref[fig:Randomcheck]{Figures~9}(a,b) show the instantaneous temperature fields of two random distributions of obstacles under $Ra=10^8$ and $\lambda=1$. Although the distribution of obstacles varies, the flow structures of both cases are similar. The dependence of heat transfer on the obstacles' thermal conductivity of two systems is illustrated in \hyperref[fig:Randomcheck]{figure~9}(c). The $Nu$ versus $\lambda$ curves based on two different obstacles' distribution show a similar trend, i.e. $Nu$ is slightly depressed as $\lambda$ increases, suggesting that the results presented in the current are robust to the random distributions of obstacles in the current parameter regime.

\end{appendices}
\clearpage

\begin{Backmatter}

\paragraph{Acknowledgements}
We acknowledge Dongpu Wang and Yuki Wakata for the insightful suggestions and discussions.

\paragraph{Funding Statement}
This work is financially supported by the National Natural Science Foundation of China under grant nos. 11988102 and 91852202, and Tencent Foundation through the XPLORER PRIZE.

\paragraph{Declaration of Interests}
The authors declare no conflict of interest.

\paragraph{Data Availability Statement}
The raw data of this study are available from the corresponding author upon reasonable request.

\end{Backmatter}
\end{document}